\documentclass[onecolumn,prd,superscriptaddress,showpacs,floatfix,%
preprintnumbers,nofootinbib,showkeys]{revtex4}  

\usepackage{axocolor}
\usepackage{amsmath,amsfonts}
\usepackage{graphicx}
\usepackage{epsfig,epsf}

\def\g {\gamma}

\def\bar {\overline}
\def\bbar {{\bar{B}}{}^0}
\def\bbbar {B^0-\bbar}

\def\be {\begin{equation}}
\def\ee {\end{equation}}
\def\beq {\begin{equation}}
\def\eeq {\end{equation}}
\def\bea {\begin{eqnarray}}
\def\eea {\end{eqnarray}}

\def\bra {\langle}
\def\ket {\rangle}

\def\bsbsbar{B_s$--$\bar{B}_s}
\def\bdbdbar{B_d$--$\bar{B}_d}

\def\dgs{\Delta \Gamma_s}

\def\g{\Gamma}

\def\beq{\begin{equation}}
\def\eeq{\end{equation}}
\def\barr{\begin{array}}
\def\earr{\end{array}}
\def\dis{\displaystyle}
 
\def\gtap{\raisebox{-.4ex}{\rlap{$\sim$}} \raisebox{.4ex}{$>$}} 

\def\opcit(#1){ {\em op. cit.}, #1}

\def\issue(#1,#2,#3){#1, #2 (#3)} 

\def\APP(#1,#2,#3){Acta Phys.\ Polon.\ \issue(#1,#2,#3)}
\def\ARNPS(#1,#2,#3){Ann.\ Rev.\ Nucl.\ Part.\ Sci.\ \issue(#1,#2,#3)}
\def\CPC(#1,#2,#3){Comp.\ Phys.\ Comm.\ \issue(#1,#2,#3)}
\def\CIP(#1,#2,#3){Comput.\ Phys.\ \issue(#1,#2,#3)}
\def\EPJC(#1,#2,#3){Eur.\ Phys.\ J.\ C\ \issue(#1,#2,#3)}
\def\EPJD(#1,#2,#3){Eur.\ Phys.\ J. Direct\ C\ \issue(#1,#2,#3)}
\def\IEEETNS(#1,#2,#3){IEEE Trans.\ Nucl.\ Sci.\ \issue(#1,#2,#3)}
\def\IJMP(#1,#2,#3){Int.\ J.\ Mod.\ Phys. \issue(#1,#2,#3)}
\def\JHEP(#1,#2,#3){J.\ High Energy Physics \issue(#1,#2,#3)}
\def\JPG(#1,#2,#3){J.\ Phys.\ G \issue(#1,#2,#3)}
\def\MPL(#1,#2,#3){Mod.\ Phys.\ Lett.\ \issue(#1,#2,#3)}
\def\NP(#1,#2,#3){Nucl.\ Phys.\ \issue(#1,#2,#3)}
\def\NIM(#1,#2,#3){Nucl.\ Instrum.\ Meth.\ \issue(#1,#2,#3)}
\def\PL(#1,#2,#3){Phys.\ Lett.\ \issue(#1,#2,#3)}
\def\PRD(#1,#2,#3){Phys.\ Rev.\ D \issue(#1,#2,#3)}
\def\PRL(#1,#2,#3){Phys.\ Rev.\ Lett.\ \issue(#1,#2,#3)}
\def\SJNP(#1,#2,#3){Sov.\ J. Nucl.\ Phys.\ \issue(#1,#2,#3)}
\def\ZPC(#1,#2,#3){Zeit.\ Phys.\ C \issue(#1,#2,#3)}

\begin{document}

\title{
B decay anomalies in an effective theory
}

\author{Debajyoti Choudhury}
\email{debajyoti.choudhury@gmail.com}
\affiliation{Department of Physics and Astrophysics, University of Delhi, 
Delhi 110007, India.}

\author{Dilip Kumar Ghosh}
\email{dilipghoshjal@gmail.com}
\affiliation{Department of Theoretical Physics, Indian Association for the 
Cultivation of Science, 2A \& 2B, Raja S.C. Mullick Road, Kolkata 700032, India. }

\author{Anirban Kundu}
\email{anirban.kundu.cu@gmail.com}
\affiliation{Department of Physics, University of Calcutta,\\
92, Acharya Prafulla Chandra Road, Kolkata 700009, India. }

\date{\today}

\begin{abstract} 

We investigate how far a new physics scenario affecting primarily the
third generation fermions can ameliorate the tension between B-decay
observables and Standard Model expectations. Adopting a model-independent 
approach, we find that among the three observables that show signs of 
such a tension, {\em viz.} the branching fractions for $B^+\to\tau\nu$, 
$B_d\to D(D^\ast)\tau\nu$, and the like-sign dimuon anomaly in neutral 
B decays, the first two can be explained adequately, while there is only 
a marginal improvement for the third. As a spin-off, it is shown that 
one can also accommodate a change in the branching fraction of the Higgs 
boson to a $\tau$ lepton pair from the SM expectation, if such a change 
is established in future data.

\end{abstract}



\maketitle

\section{Introduction}

While the purported discovery of the Higgs boson at the Large Hadron
Collider (LHC) \cite{cms-higgs,atlas-higgs} seems to vindicate the
Standard Model (SM), there are enough reasons to believe that the
latter is but an effective theory, valid only up to a certain energy
scale, with a more complete theory lurking beyond that. One of the
major reasons for such a belief is the fine-tuning problem associated
with such an elementary scalar; also of considerable import are issues
such as the existence of the dark matter, or the baryon asymmetry in
the universe.

This acts as a strong motivation to look for signals, both direct and
indirect, of such a new theory. While direct signals very often
involve production of new particles, the indirect signals will, most
probably, be manifested as modifications to SM observables by new
effective operators, or even the same operators as in the SM,
but with modified Wilson coefficients.

B meson observables have constituted a favourite
hunting ground for indirect signals. Over the years, several
experiments, including B-factories, Tevatron, and even the LHC, have
reported observables that are not in good agreement with the SM. While
the tension is not so overwhelming as to claim unquestioned evidence
of New Physics (NP), the pattern is interesting. Here, one must
remember to tackle the theoretical uncertainties carefully; some of
the discrepancies, like the longitudinal polarization
anomaly in the decay of a B meson to two vector mesons,
vanished because of a more careful reappraisal of the SM effects.

Let us begin by considering a few observables which are not in full
conformity with the SM expectations:
\begin{itemize}
 
 \item the large branching ratio of $B \to D(D^\ast)\tau\nu$, 
with a combined tension of $3.4\sigma$~\cite{babar-dtaunu};
 \item the large branching ratio of $B^+\to \tau^+\nu$, 
with a tension of $1.6\sigma$~\cite{ckmfitter-eps11}
\footnote{The tension has come down very recently; it was about $2.8\sigma$
before the publication of the latest Belle result \cite{belle-btaunu}.};
 \item the like-sign dimuon asymmetry, with a tension of $3.9\sigma$~\cite{d0-dimuon}.
\end{itemize}
It is interesting to note that the first two involve a $\tau$ lepton
in the final state. This motivates us to ask if there exists one or
more new effective operators involving the $b$ quark and the $\tau$
lepton. Such a possibility was raised in Ref.\ \cite{dkn1}, and
further investigated in
Refs.~\cite{dkn2,bauer,dgkp,bobeth,datta,dg}.  One might feel
tempted to add to this list a hint of another anomaly: the branching
ratio of $H\to\tau^+\tau^-$ seems to be a bit on the lower side than that
expected in the SM \cite{cms-higgs}.

The lowest dimensional operators of interest can, generically, be
expressed as $(\bar{b}\, \Gamma_A \, s) \, (\bar\tau \, \Gamma_B \,
\tau)$ where $\Gamma_{A,B}$ are appropriate combinations of the Dirac
matrices. As shown in Ref.~\cite{bauer}, this four-fermion operator is
relatively unconstrained. This leads to a new contribution to the
$\bsbsbar$ mixing amplitude, with a nonzero absorptive part: cutting
the intermediate $\tau$ propagators can yield an on-shell $\tau$-pair.
Thus, one has a new contribution to $\dgs$, the difference in the
widths of the two $B_s$ mass eigenstates, which, in turn, ameliorates
the apparent discrepancy in the like-sign dimuon asymmetry.  However,
the strength of any such operator is ultimately constrained by the
mass difference $\Delta M_s$ of the $B_s$ mass eigenstates.

Considering the fact that there is hardly any tension in the data
involving electrons or muons in the final state (except the dimuon
anomaly, to which we will come shortly), one might feel tempted to
invoke one or more effective operators involving only third generation
quarks and leptons.  While such an effective operator based study was
undertaken in Refs.~\cite{dgkp,dg}, the constraints on $\Delta M_s$
were not correlated with those coming from the decay width difference
$\dgs$; they were assumed to be independent numbers. The authors of
Ref.~\cite{bobeth} discussed the effectiveness of $\Delta M_s$ as a
possible constraint on the parameter space.

In this paper, we adopt a different approach. To begin with, we posit
a single effective operator involving a third-generation quark current
and a third-generation lepton current.  As it involves only
third-generation fields, the constraints on the Lorentz structure for
the same, or on the magnitude of the corresponding Wilson coefficient
is relatively weak. For example, just restricting the new couplings to
the perturbative regime ensures that $\Upsilon(1S)\to\tau^+\tau^-$
does not receive a significant contribution over and above the SM
amplitude, which is electromagnetic in nature.  We will, however, not
venture to discuss any particular models that might predict such an
interaction, and adopt, instead, a bottom-up approach.

A theory of flavour would, generically, dictate that such an operator
would be written in the weak basis. On the breaking of the electroweak
symmetry, the quark fields would need to be rotated to the mass basis.
This leads to a plethora of new operators, related to the original
through the Cabibbo-Kobayashi-Maskawa (CKM) matrix elements. Included,
amongst others, are those leading to $b\to s\tau^+\tau^-$ and $b\to
s\nu\bar\nu$ (the lepton fields are not rotated, so we will always
write $\nu$ for $\nu_\tau$).  We will see that, even with a moderate
NP scale $\sim 1$ TeV, the constraints from $\Delta M_s$ are so strong
for vector and axial-vector $(\{V,A\})$ operators that the effects on
$\dgs$ and $B_s\to\tau^+\tau^-$ are bound to be unobservably
small. Thus, the explanation of the dimuon anomaly must lie elsewhere,
while this scheme can successfully explain the charged-current
decays. The outlook is better if the effective operators are of scalar
and pseudoscalar $(\{S,P\})$ variety.  Indeed, such a scenario
predicts a rather strong enhancement of the branching fraction of
$B_c\to\tau\nu$ over its SM prediction.  As for the tensor current
operators, the corresponding Wilson coefficients are severely
constrained \cite{bobeth} from radiative decays like $b\to s\gamma$,
and so we will not consider them any further.

The rest of the paper is arranged as follows. In the next section, we will briefly
go through the existing data. In Section III, we will discuss the new
operators; first, the $\{S,P\}$ type, and then the $\{V,A\}$ type. 
In Section IV, we show how these
operators may help in bringing down the tension with the SM. We
summarize and conclude in Section V.
\section{Existing constraints}

\subsection{$B\to D(D^\ast) \tau\nu$}

The importance of studying the $B\to D(D^\ast)\tau \nu$ modes for a
possible signal of new physics has already been pointed out in the
literature \cite{nierste-etal}.  The BaBar Collaboration
\cite{babar-dtaunu} measured the branching fractions for these two
modes, and they are above the SM predictions.  They are also not
consistent with a type II two-Higgs doublet model (such as the 
minimal supersymmetric extension of the SM). 
The implications of the data as a possible hint of physics beyond the
SM have been studied in \cite{datta}.

It is particularly useful to consider the
ratios $R(D)$ and $R(D^\ast)$, defined as
\beq
R(D^{(*)}) = \frac{{\rm Br}(B\to D^{(*)}\tau\nu)} 
{{\rm Br}(B\to D^{(*)}\ell\nu)}
\eeq
as these are largely free of the uncertainties---{\em e.g.}, those 
in the form factors---that exclusive modes are often prey to.
The SM predictions are
\beq
R(D) = 0.297 \pm 0.017\,, \qquad R(D^*) = 0.252\pm 0.003\,,
\label{dtaunu:1}
\eeq
while the BaBar Collaboration quotes \cite{babar-dtaunu}
\beq
R(D) = 0.440 \pm 0.058 \pm 0.042\,, \qquad
R(D^*) = 0.332 \pm 0.024 \pm 0.018\,.
\label{dtaunu:2}
\eeq
It should be noted that a recent calculation \cite{lattice-dtaunu}
in unquenched lattice QCD gives,
 in the SM, $R(D) = 0.316 \pm 0.012
\pm 0.007$. This is consistent with the earlier SM prediction, 
but cannot explain the tension with the 
data\footnote{The Belle collaboration measurements\cite{belle-dtaunu}
 viz. 
\[
R(D) = 0.70^{+0.19+0.11}_{-0.18-0.09}\,, \qquad
R(D^*) = 0.47^{+0.11+0.06}_{-0.10-0.07}\,.
\]
while being even further away from the SM expectations, nonetheless 
are consistent with these as well as the BaBar results. This agreement, 
though, is but a consequence of  the large error margins.}.

Using Eqs.\ (\ref{dtaunu:1}) and (\ref{dtaunu:2}), and adding all errors
in quadrature, we get
\beq
\frac{R(D)_{exp}}{R(D)_{SM}} = 1.481 \times (1\pm 0.173)\,,\ \ 
\frac{R(D^\ast)_{exp}}{R(D^\ast)_{SM}} = 1.317 \times (1\pm 0.091)\,. 
\eeq

\subsection{$B\to\tau\nu$ and $B_c\to\tau\nu$}

The partial decay width  $B\to\tau\nu$, in the SM, is given by 
\beq
\Gamma (B\to\tau\nu) = \frac{1}{8\pi} G_F^2 \vert V_{ub}\vert^2 f_B^2 m_\tau^2 m_B \left(
1- \frac{m_\tau^2}{m_B^2}\right)^2\,.
\label{btaunu}
\eeq
The world average is \cite{ckmfitter-eps11} 
\beq
{\rm Br}(B\to\tau\nu) = (11.5\pm 2.3)\times 10^{-5}\,,
\eeq
while the theoretical prediction is 
\beq
{\rm Br}(B\to\tau\nu)_{\rm SM} = \left(7.57^{+0.98}_{-0.61}\right)\times 10^{-5}\,,
\eeq
which gives a tension at the level of $1.6\sigma$ \cite{ckmfitter-eps11}.
The theoretical
uncertainty comes from those in 
the decay constant $f_B$ and the CKM matrix element $V_{ub}$. 
While the discrepancy has eased considerably, from $2.8\sigma$ to $1.6\sigma$, 
after the publication of the new 
Belle result \cite{belle-btaunu}, there is still a non-negligible tension between 
the value of $|V_{ub}|$ determined from this decay, and that determined
indirectly from the sides of the unitarity triangle, or an average of direct inclusive 
($B\to X_u\ell\nu$) and exclusive ($B\to\pi\ell\nu$) measurements.

The discrepancy has led to several attempts in the literature to explain this 
as a possible NP signal. However, the explanations based on the existence of 
only a charged Higgs boson of type-II are ruled out at 95\% confidence limit 
from the combined analysis of processes like $b\to s\gamma$, $Z\to b\bar{b}$, 
$\bbbar$ mixing \cite{deschamps}; the goodness-of-fit is hardly better than the 
fit with the SM alone. Models like R-parity violating supersymmetry 
fare much better and give a satisfactory explanation of the excess
\cite{bose}. Note that all those analyses were performed using the pre-2012 data.

A similar expression as in Eq.\ (\ref{btaunu}) holds for $B_c \to\tau\nu$.
For numerical evaluation, one might use
\beq
f_{B_c} = (395\pm 15)~{\rm MeV}\,,\ \ \tau_{B_c} = 0.458\pm 0.030~{\rm ps}\,.
\eeq

\subsection{$\bsbsbar$ mixing: $\Delta M_s$, $\Delta \Gamma_s$, $\beta_s$ and $\phi_s$}

While there are no apparent tensions in this sector at present, the data, 
as we will soon see, acts as a very tight constraint on NP operators.
The mass splitting between two $B_s$ mass eigenstates, $\Delta M_s \approx 2|M_{12s}|$, 
is extremely well-measured \cite{hfag}, namely
\beq
\Delta M_s = 17.719\pm 0.043 ~{\rm ps}^{-1}\,. 
\eeq
This agrees very well with the SM expectation \cite{lenz}, {\em viz.}
\beq
\Delta M_s~({\rm SM}) = (17.3\pm 2.6)~{\rm ps}^{-1}\,,
\eeq
and acts as a very tight constraint on NP models.
There are two relevant phases in the $\bsbsbar$ system. The first one, the mixing phase, is defined as
\beq
\beta_s = {\rm arg}\left(- \, \frac{V_{cb}V_{cs}^\ast}{V_{tb}V_{ts}^\ast}\right)\,,
\eeq
while the second one, responsible for semileptonic asymmetries, is given by
\beq
\phi_s = {\rm arg}\left(-\frac{M_{12s}}{\Gamma_{12s}}\right)\,.
\eeq
The SM predictions~\cite{hfag} are
\beq
\phi_s = 0.0041\pm 0.0007\,,\  -2\beta_s = -0.038 \pm 0.002\,.
\eeq
The experimental numbers are 
\beq
-2\beta_s = -0.040 ^{+0.090}_{-0.085}
\eeq
from direct determination, and 
\beq
-2\beta_s = -0.0363^{+0.0016}_{-0.0015}
\eeq
from an indirect global fit \cite{hfag}. 
We will use the former number. 

Note that $B_s$ lifetime is rather ill-defined, as the two mass
eigenstates have a significant lifetime difference, namely
 \beq
 \tau_{B_{sL}} = 1.408 \pm 0.017~{\rm ps} \ , \qquad
\tau_{B_{sH}} = 1.626 \pm 0.023~{\rm ps} \,.
\eeq
Averaging over the two, we have
\beq
 \tau_{B_s} ({\rm average}) = \frac{2}{\Gamma_L+\Gamma_H} = 1.509 \pm 0.012~{\rm ps}\,,
\eeq
which should be compared with $\tau_{B_d} = 1.519 \pm 0.007$ ps. Thus, 
\beq
\frac{\tau_{B_s}}{\tau_{B_d}} = 0.993\pm 0.009 \ ,
\eeq
while the SM expectation for this ratio 
lies between 0.99 and 1.01 \cite{hfag}.

The width difference $\dgs$ is given by
\beq
\dgs = 2 \, \vert \Gamma_{12s}\vert \cos(\phi_s)\,.
\eeq
While the SM predicts $\dgs > 0$, there was a sign ambiguity earlier in its determination.
Recently LHCb, from the decay $B_s\to J/\psi K^+K^-$, found $\dgs > 0$ 
with a $4.7\sigma$ confidence level
\cite{lhcb-signgamma}.
The experimental number, an average over various measurements
\cite{hfag},
\beq
\dgs = 0.095\pm 0.014~{\rm ps}^{-1}
\eeq
is to be compared with the SM prediction \cite{lenz}
\beq
\dgs~({\rm SM}) = 0.087\pm 0.021~{\rm ps}^{-1}\,.
\eeq

\subsection{The like-sign dimuon asymmetry}

The like-sign dimuon asymmetry, 
defined as
\beq
A^b_{sl} = \frac{N(\mu^+\mu^+)-N(\mu^-\mu^-)}{N(\mu^+\mu^+)+N(\mu^-\mu^-)}\,,
\eeq
and measured with 
9.0 fb$^{-1}$ of data at the D\O~Collaboration is \cite{d0-dimuon}
\beq
A^b_{sl} = (-7.87\pm 1.96)\times 10^{-3}\,.
\label{Absl_exp}
\eeq
This can be expressed as individual flavour-specific (fs) semileptonic asymmetries coming from $B_d$ and $B_s$:
\beq
A^b_{sl} = (0.595\pm 0.022)\, a^d_{fs} + (0.405\mp 0.022) \, a^s_{fs}\,,
\eeq
where the numbers in the parentheses are the production fractions for $B_d$ and $B_s$.
The SM expectations are
\beq
a^d_{fs} = (-4.1\pm 0.6)\times 10^{-4}\,,\ \ a^s_{fs} = (1.9\pm 0.3)\times 10^{-5}\,,
\eeq
which give the SM prediction
\beq
(A^b_{sl})_{SM} = (-2.4\pm 0.4)\times 10^{-4}\,.
\label{Absl_theo}
\eeq
Comparing Eqs.~(\ref{Absl_exp}) and (\ref{Absl_theo}), one finds a $3.9\sigma$ discrepancy 
between theoretical prediction and experiment.
$a^d_{fs}$ has already been measured by BaBar and Belle; the 
average \cite{hfag}
\beq
a^d_{fs} = (-3.3\pm 3.3)\times 10^{-3}
\eeq
is consistent with the SM. This gives an indirect prediction for
$a^s_{fs}$, {\em viz.}
\beq
a^s_{fs} = (-1.52\pm 1.04)\times 10^{-2}\,,
\eeq
where the error has been symmetrized. 
We have neglected the correlation between $a^d_{fs}$ and $a^s_{fs}$, but
have taken the uncertainties in the production fractions into account. 
Recently, the D\O~Collaboration directly measured 
$a^s_{fs} = (-1.08 \pm 0.72~({\rm stat}) \pm 0.17~({\rm syst}))\times 10^{-2}$
\cite{d0-asfs} which is also consistent with the SM expectation.  
The HFAG collaboration
averages over several direct measurements of $a^s_{fs}$ and quotes \cite{hfag}
\beq
a^s_{fs} = (-1.05 \pm 0.64) \times 10^{-2}
\eeq
but this has a nonzero correlation with $a^d_{fs}$.

This gives a weak constraint on $\phi_s$:
\beq
\tan\phi_s = a^s_{fs} \, \frac{\Delta M_s}{\Delta\Gamma_s}\,.
\eeq
If there is some NP contributing to both $M_{12s}$ and 
$\Gamma_{12s}$, one can parametrize the NP contribution as
\beq
\begin{array}{rclcl}
M_{12} &=& M_{12}^{SM} + M_{12}^{NP} & \equiv &  M_{12}^{SM} \, R_M \, 
\exp(i\phi_M)\, ,\\[1ex]
\g_{12} &=& \g_{12}^{SM} + \g_{12}^{NP}  & \equiv &  \g_{12}^{SM} \, R_{\g} \, 
   \exp(i\phi_\g)\, \ ,
\end{array}
\eeq
resulting in \cite{dkn2}
\beq
\phi_s = \phi_s^{SM} + \phi_M - \phi_\Gamma\,.
\eeq

Thus, there are two ways to have a large $a^s_{fs}$; either a large
contribution to $\dgs$ or a large $\phi_s \sim \pi/2$. But
$\phi_s^{SM}\approx 0$ and $\phi_M \equiv -2\beta_s$ is known to be
small, so a large $\phi_s$ necessarily warrants a large $\phi_\Gamma$,
and hence a large contribution to $\g_{12}$.

Taking all the existing constraints into account, it was shown 
\cite{dkn1,dkn2,bobeth} that $b\to s\tau^+\tau^-$ is a viable
option to generate a large $\g_{12}$. 
However, such a new channel decreases the lifetime of $B_s$
compared to $B_d$; moreover, one does not expect ${\rm
  Br}(B_s\to\tau^+\tau^-)$ to be more than 3-$3.5$\%
\cite{bobeth}. The inclusive mode $B(B_d\to X_s\tau\tau)$ is
constrained to be less than 5\% \cite{aleph-bstautau}, while BaBar
gives a 90\% limit \cite{babar-bstautau}
\beq
B(B^+\to K^+\tau^+\tau^-)\vert_{q^2 > 14.23~{\rm GeV}^2} < 3.3\times 10^{-3}\,.
\eeq  
However, not all Lorentz structures that contribute to a new
absorptive part in $\bsbsbar$ mixing contribute simultaneously to
$B_s\to \tau^+\tau^-$ or $B^+\to K^+\tau^+\tau^-$.  At the same time,
there can be significant long-distance effects in $\bsbsbar$ mixing,
through meson loops, and they can have a non-negligible contribution in
$\dgs$ \cite{aleksan,chua}. 
While we will discuss these issues in detail
later, the crucial point to note is that the NP contributing to $\g_{12}$ 
should, in general, contribute to $M_{12}$ also, and the mass difference
$\Delta M_s$ is so tightly constrained that this leaves only a very small
room for any NP.

\subsection{Di-tau suppression}

Looking for the SM Higgs in the $H\to \tau\tau$ mode, the CMS 
collaboration has failed to see \cite{cms-higgs}
an unambiguous excess over the background.
Indeed, for the preferred mass of $m_H=125$ GeV (corresponding to the 
much-touted diphoton and four-lepton excesses), it is only able to 
impose a 95\% C.L. upper limit on the ditau excess.
With the result being similar for the ATLAS 
collaboration \cite{atlas-higgs} as well,  
the $p p \to H\to\tau\tau$ 
cross-section is, in fact, consistent with zero, 
viz., $\sigma/\sigma_{SM} = 0.100^{+0.714}_{-0.699}$ \cite{Eberhardt:2012gv}.

\begin{table}[htbp]
\begin{center}
\begin{tabular}{||c|c|c||}
\hline
Observable & SM & Expt \\
\hline
$-2\beta_s$ & $-0.038\pm 0.002$  & $ -0.040 ^{+0.090}_{-0.085}$    \\
$\tau_{B_s}$ &    & $1.509\pm 0.012$ ps  \\
$\tau_{B_s}/\tau_{B_d}$ & 0.99 - 1.01 & $0.993\pm 0.009$   \\
$\Delta M_s\approx 2|M_{12s}|$ & $(17.3\pm 2.6)$ ps$^{-1}$ & $(17.719\pm 0.043)$ ps$^{-1}$ \\
$\Delta \Gamma_s\approx 2|\Gamma_{12s}|\cos\phi_s$ & $(0.087\pm 0.021)$ ps$^{-1}$ & $(0.095\pm 0.014)$ ps$^{-1}$ \\
$A^b_{sl}$ & $(-2.4\pm 0.4)\times 10^{-4}$ & $(-7.87 \pm 1.96)\times 10^{-3}$ \\
$a^d_{fs}\equiv a^d_{sl}$ & $(-4.1\pm 0.6)\times 10^{-4}$ & $(-3.3\pm 3.3)\times 10^{-3}$  \\
$a^s_{fs}\equiv a^s_{sl}$ & $(1.9\pm 0.3)\times 10^{-5}$ & $(-1.05\pm 0.64)\times 10^{-2}$  \\
\hline
$R(D)$ & $0.297\pm 0.017$ & $0.440\pm 0.072$ \\
$R(D^\ast)$ & $0.252\pm 0.003$ & $0.332\pm 0.030$ \\
Br($B^+\to\tau^+\nu$) & $(7.57^{+0.98}_{-0.61})\times 10^{-5}$ & $(11.5\pm 2.3)\times 10^{-5}$ 
\\
\hline  
\end{tabular}
\caption{Inputs used in our analysis. For details, see text.}
\end{center}
\end{table}

\subsection{Numbers used for the analysis}

Apart from the numbers shown in the previous subsections, 
a summary of which is given in Table 1,
we also use the following for our analysis:
\beq
m_{B^+} = m_{B_d} = 5.279~{\rm GeV}\,, \quad 
\tau_{B^+} = 1.641~{\rm ps}\,, \quad
\tau_{B_d} = 1.519~{\rm ps}\,, \quad 
m_{B_s}=5.367~{\rm GeV}\,,
\eeq
and
\bea
&{}&|V_{td}| =  \dis (8.67^{+0.29}_{-0.31})\times 10^{-3}\,, \ \ 
|V_{ts}| =  \dis 0.0404^{+0.0011}_{-0.0005}\,,\ \ 
|V_{cb}| = \dis 0.0412^{+0.0011}_{-0.0005}\,,\nonumber\\
&{}& |V_{ub}| = (3.49\pm 0.13)\times 10^{-3}\,,\ \ 
\gamma \approx \dis {\rm arg}(V_{ub}^\ast) = 77^\circ\,.
\eea
Note that $|V_{ts}|$ is measured from $\bsbsbar$ mixing, but if we
talk about new physics in the mixing and hence $\Delta M_s$, we should, 
instead, 
use $|V_{ts}|$ as determined from the unitarity constraints.  The
central value as determined from the unitarity is $0.0404$; purely
from $\Delta M_s$ measurement, this comes out to be $0.0429 \pm
0.0026$. 
The error margin in $\gamma$ is not important for our
analysis, and so we use the central value \cite{hfag}. Note that 
only the difference of $\gamma$ and the weak phase coming in the 
NP amplitude is relevant for our purpose; the latter is 
{\em a priori} unknown and must be treated as a free parameter of the theory.
For the
evaluation of $\Delta M_s$, we have used the unquenched lattice value
\beq
f_{B_s} \sqrt{B_{B_s}} = 248 \pm 15~{\rm MeV}\,.
\eeq

\section{The new effective operators}

Let us, now, consider a set of possible operators involving
third-generation fermions, satisfying both Lorentz and gauge
invariance. These might be of the form 
\beq
{\cal O}_S =
A(\bar{Q}_{3L} d_{3R})(\bar{e}_{3R}L_{3L}) + B(\bar{Q}_{3L}u_{3R})(\bar{e^c}_{3R} {L'}^c_{3L}) + {\rm h.c.}
\label{op-sp}
\eeq
with ${L'}_3 \equiv i\sigma_2 L_3$, or
\beq
{\cal O}_V = 
C \left[ \bar{Q}_{3L} \gamma^\mu \tau_a Q_{3L}\right] \left[ \bar{L}_{3L} \gamma_\mu \tau_a L_{3L}\right]\,,
\label{op-va}
\eeq
where $Q_3$, $L_3$, $u_3$, $d_3$, and $e_3$ stand for the doublet
quark, doublet lepton, singlet up-type, singlet down-type, and singlet
charged lepton of the third generation respectively.  In view of
  the experimental measurements that we seek to address, we limit
  ourselves, here, to only those operators that admit charged-current
  interactions. Furthermore, we do not consider tensor operators as
  their Wilson coefficients are very tightly constrained from
  radiative decays.  In terms of component fields, we can write
the scalar-pseudoscalar operators as
\bea
{\cal O}_S&=& 
A\left[(\bar{b}P_R b)(\bar\tau P_L \tau) + (\bar{t} P_R b)(\bar\tau P_L \nu) + 
(\bar{b}P_L b)(\bar\tau P_R \tau) + (\bar{b} P_L t)(\bar\nu P_R \tau)\right] \nonumber\\
&& + 
B\left[(\bar{t}P_R t)(\bar\tau P_L \tau) - (\bar{b} P_L t)(\bar\nu P_R\tau) +
(\bar{t}P_L t)(\bar\tau P_R \tau) - (\bar{t} P_R b)(\bar\tau P_L \nu)\right]\,, \nonumber\\
&=& A\left[ (\bar{b}P_R b)(\bar\tau P_L \tau) + {\rm h.c.}\right]
+ B\left[ (\bar{t}P_R t)(\bar\tau P_L \tau) + {\rm h.c.}\right] \nonumber\\
&& + (A-B) \left[ (\bar{t} P_R b)(\bar\tau P_L \nu) + (\bar{b} P_L t)(\bar\nu P_R \tau)\right]\nonumber\\
&=& \frac12 A \left[(\bar{b}b)(\bar\tau\tau) - (\bar{b}\gamma_5 b)(\bar\tau\gamma_5\tau)\right]
+{\rm similar~terms}\,.
\label{op-sp2}
\eea
Eq.\ (\ref{op-sp2}) shows that the neutral current operators 
have a coefficient different from that for
 the charged current operators. In fact,
there are two neutral current operators now, one involving scalar
currents and the other involving pseudoscalar currents. While we will
discuss later the consequences of such an operator structure, note
that the $\tau$ Yukawa coupling can, in principle, be significantly
modified by a top loop. Corrections to the Yukawa couplings of other
third generation fermions are negligible.

In a similar vein,
\bea
{\cal O}_V&=& C \left[ (\bar{b}\gamma^\mu P_L t)(\bar\nu\gamma_\mu P_L \tau) +
(\bar{t} \gamma^\mu P_L b)(\bar\tau\gamma_\mu P_L \nu) \right. \nonumber\\
&& 
+ \frac12 \left( \bar{b}\gamma^\mu P_L b -  \bar{t} \gamma^\mu P_L t)
(\bar\tau\gamma_\mu P_L \tau - \bar\nu\gamma_\mu P_L \nu) \right]\,.
\label{op-va2}
\eea
To make explicit the higher-dimensional nature of the couplings, we denote
\beq
A = a/\Lambda^2\,,\ \ \ B = b/\Lambda^2\,,\ \ \ C = c/\Lambda^2\,,
\eeq
where $a$, $b$, and $c$ are dimensionless couplings.

If Eqs.\ (\ref{op-sp2}) and/or (\ref{op-va2}) are all we have, the
phenomenology is straightforward, and only a 
subset of that we would consider below.  One might think that
$\Upsilon\to\tau\tau$ will put a tight constraint on the coefficients,
but, in actuality, 
that constraint is far too weak. The reason is that the SM decay is
an electromagnetic one, and the width is given by
\cite{sanchis-lozano}
\beq
\Gamma_{\Upsilon(1S)\to \ell\ell} = 4\, \alpha^2 \, Q_b^2 \, 
M_{\Upsilon}^{-2} \, |R(0)|^2 \, 
(1+2x) \, \sqrt{1-4x} \,,
\eeq
where $x = M_\ell^2/M_\Upsilon^2$ and $R(0)$ is the radial part of the
non-relativistic wave function at the origin.

\subsection{The rotation}

Assuming that
  the operators in question have arisen on account of some flavour
  physics operative at scales higher than the weak scale, we now put
forward the Ansatz that the fields in Eqs. (\ref{op-sp2}) and
(\ref{op-va2}) are in the weak basis, and should be rotated to the
stationary or mass basis.  Let us, for simplicity, assume that
right-chiral fields are not rotated, and for the left-chiral fields,
\beq
b_{wk} \to x_1 d + x_2 s + x_3 b\,,\ \ t_{wk} \to y_1 u + y_2 c + y_3 t\,,
\eeq
where the right-hand side fields are in the mass basis.

If ${\cal U}$ and ${\cal D}$ matrices are responsible for the rotation
of $T_3=+1/2$ and $T_3=-1/2$ fields from the weak basis to the mass
basis, so that the CKM matrix $V = {\cal U}^\dag {\cal D}$, one notes
that $(x_1,x_2,x_3)$ and $(y_1,y_2,y_3)$ are just the third rows of
${\cal D}$ and ${\cal U}$ respectively. If we assume the rotation
matrices to be almost diagonal, the only constraint is
\beq
y_3^\ast x_3 \approx V_{tb}\,.
\eeq
As for other combinations, we can, at most,
use order-of-magnitude arguments to yield
\beq
y_3^\ast x_1 \sim V_{td}\,,\  y_3^\ast x_2 \sim V_{ts}\,,\  y_2^\ast x_3 \sim V_{cb}\,,\  y_1^\ast x_3 \sim V_{ub}\,,
\eeq
although there can be significant deviations. 
Note that this is a rather conservative constraint and one can build models 
to bypass this. However, one has to be extremely careful about
constraints coming from flavour physics, in particular those involving 
fermions of the first two generations. Furthermore, 
such models involve some degree of
fine-tuning between the rotations in the right-chiral and left-chiral quark
sectors.

\section{The observables}

\subsection{Leptonic and semileptonic decay channels: $B^+\to \tau^+\nu$, $B_c\to \tau^+\nu$, $B \to D(D^\ast) \tau\nu$} 

The relevant new operators are
\bea
{\cal O}_S &\supset& \frac{1}{\Lambda^2}(a-b)\left[
\left( \{y_1^\ast \bar{u} + y_2^\ast \bar{c}\} P_R b \right) \; (\bar\tau P_L\nu) + {\rm h.c.}\right]\,,\nonumber\\
{\cal O}_V &\supset& \frac{1}{\Lambda^2} c  \, x_3\left[
\left(\{y_1^\ast \bar{u} + y_2^\ast \bar{c}\}) P_L b\right) \; (\bar\tau P_L\nu) + {\rm h.c.}\right]\ , 
\eea
and their effect on the amplitudes of interest
can be obtained by simple replacements in the corresponding 
SM expressions, namely,
\beq
\begin{array}{rclcl}
\displaystyle \frac{G_F}{\sqrt{2}} V_{ub} &\to& 
\displaystyle \frac{G_F}{\sqrt{2}} V_{ub} + \frac14 \frac{c}{\Lambda^2} y_1^\ast x_3
& \qquad & {\rm for}~{\cal O}_V\,,\\[2ex]
\displaystyle \frac{G_F}{\sqrt{2}} V_{ub} m_\tau &\to& 
\displaystyle \frac{G_F}{\sqrt{2}} V_{ub} m_\tau - \frac14
\frac{a-b}{\Lambda^2} y_1^\ast \frac{m_{B}^2} {m_b+m_u} & & 
{\rm for}~{\cal O}_S\,,
\end{array}
\eeq
where the latter follows from 
 \beq
\bra 0 | \bar{u} (1-\gamma_5) b|B^-\ket = -if_{B}\frac{m_{B}^2}{m_b+m_u}\,.
\eeq   
For the $B_c$ decay, one has to make the following substitutions: $\{u,V_{ub},y_1,m_B,m_u,f_B\}
\to \{c,V_{cb},y_2,m_{B_c},m_c,f_{B_c}\}$.

\begin{figure}
\vspace{-10pt}
\centerline{
\rotatebox{0}{\epsfxsize=8cm\epsfbox{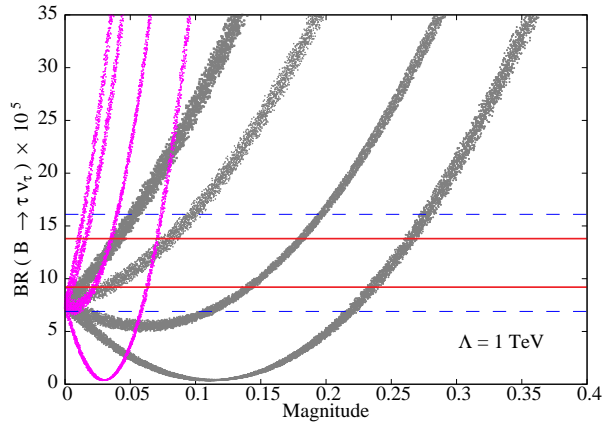}}
}
\caption{The dependence of ${\rm Br}(B \to \tau \nu)$ on the magnitude 
  of the new physics couplings $\vert c y_1^\ast
  x_3\vert$ (thick black bands) and $\vert (a-b) y_1^\ast\vert$ (thin purple/light grey bands). In each case, the different 
  bands from left to right correspond to differing values of the phase 
of the new physics coupling, 
namely, $-\pi/2, 0, \pi/4 $ and $\pi/2$ respectively  
   and the thickness of the individual 
  band reflects the errors in $|V_{ub}|$ and $f_B$ at the $1\sigma$ level. 
  The experimental data on
  ${\rm Br} (B\to\tau\nu)$ at $1\sigma$ (red/dark grey solid lines) and $2\sigma$ (blue/light grey broken lines) 
intervals are shown as horizontal bands.}
   \label{fig:btaunu}
\end{figure}
Taking only one set of new physics couplings, $c x_3 y^\ast_1$ 
or $(a-b) y^\ast_1$, to be non-zero, in
Fig.~\ref{fig:btaunu} we show  the variation of 
${\rm Br}(B \to \tau \nu) $ with the magnitude of the coupling. 
We have
set the scale of the new physics $\Lambda = 1 $ TeV, and used
$\phi_3 = \gamma = 77^\circ$ \cite{hfag}.
Understandably, the phase of the NP coupling plays a significant 
role with positive values allowing for destructive interference 
with the SM amplitude. This results in the different bands (one for 
each representative value of the phase).
The width of the bands is due to the uncertainty in $V_{ub}$ and to a
lesser extent, that in $f_B$.
The two horizontal bands correspond to $1\sigma $ (red/dark grey lines) and 
$2\sigma $ (blue/light grey lines) intervals of 
experimental data on ${\rm Br}(B \to \tau \nu )$.
Their intersection with the NP bands determine 
the allowed ranges for the couplings.
Note that if $|y_1|$ is 
indeed ${\cal O}(|V_{ub}|)$, 
then $|cy_1^\ast x_3| > {\cal O}(0.1)$ 
would indicate a significant departure from
the expectations in naive dimensional analysis.


For $B\to D(D^\ast)\tau\nu$ and $B_c\to\tau\nu$, the SM effective Lagrangian is
\beq
{\cal L}_{eff} =  \frac{4G_F V_{cb}}{\sqrt{2}} \left(\bar{c}\gamma^\mu P_L b\right) 
\left(\bar\tau\gamma_\mu P_L\nu\right)\,.
\eeq
The two NP operators ${\cal O}_S$ and ${\cal O}_V$ modify this to
\beq
{\cal L}_{NP} =  \frac{4G_F V_{cb}}{\sqrt{2}} \left[ (1+g') \left(\bar{c}\gamma^\mu P_L b\right)
\left(\bar\tau\gamma_\mu P_L\nu\right)
+ g_R \left(\bar{c} P_R b\right)\left(\bar\tau P_L\nu\right)
\right]\,, 
\eeq
where the $g'$ ($g_R$) term emanates from ${\cal O}_V$ (${\cal O}_S$), 
namely
\beq
\frac{4G_F V_{cb}}{\sqrt{2}} g' = C y_2^\ast x_3\,,\ \ 
\frac{4G_F V_{cb}}{\sqrt{2}} g_R = (A-B) y_2^\ast\,. 
\eeq
If $g'\not = 0$ but $g_R=0$, one can write
\beq
R(D) = R_{SM}(D) \, \vert 1+g'\vert^2
\eeq
and a similar equation for $R(D^*)$, assuming that the new interaction
does not contribute to the electron or muon channel.
On the other hand, if $g_R \not=0$ but $g'=0$, we get
\bea
R(D) &=& R_{SM}(D) \left[ 1+ 1.5 \, {\rm Re}(g_R) + \vert g_R\vert^2 \, \right]
\,,\nonumber\\
R(D^\ast) &=& R_{SM}(D^\ast) \left[ 1+ 0.12 \, {\rm Re}(g_R) 
            + 0.05 \, \vert g_R\vert^2 \, \right]\,.
\eea
The same couplings also contribute to the 
leptonic decay $B_c\to\tau\nu$, and depending on
the phase of the coupling, can increase or decrease the branching ratio. 

\begin{figure}
\vspace{-50pt}
\centerline{
\rotatebox{0}{\epsfxsize=8cm\epsfbox{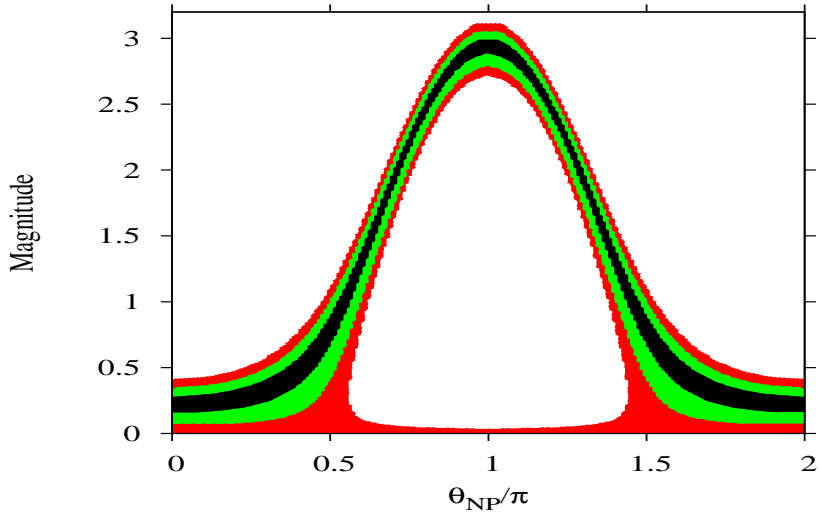}}
}
\vspace*{-2cm}
\caption{Allowed values of the magnitude and phase of the new physics
  coupling $c y_2^\ast x_3$.  The black, green (light grey), and red
  (dark grey) regions denote the $1\sigma$, $2\sigma$, and $3\sigma$
  consistency bands respectively, where both theoretical and
  experimental errors are taken into account and added in quadrature.}
   \label{fig:bdtaunu}
\end{figure}

In Figure \ref{fig:bdtaunu}, we show the allowed values of the
coupling $cy_2^\ast x_3$, with $\Lambda = 1$ TeV, at different
confidence levels. 
The intervals are calculated with individual error margins 
and not with a combined $\chi^2$-fit.
Corresponding to these three cases, the branching
ratio of $B_c\to\tau\nu$ are
\beq
{\rm Br}(B_c\to\tau\nu) \in [2.05-2.40]\%~~(1\sigma)\,,\ \ 
[1.80 - 2.60]\%~~(2\sigma)\,,\ \ 
[1.60 - 2.80]\%~~(3\sigma)\,,
\eeq
which should be compared with the SM value of $1.68\%$.

For the $\{S,P\}$ couplings, there is no region in the parameter space
compatible with both $R(D)$ and $R(D^\ast)$ at $1\sigma$ or $2\sigma$
level. This is because of the small contribution of $g_R$ to
$R(D^\ast)$. Only at $3\sigma$, does one get an allowed region in the
parameter space. However, ${\rm Br}(B_c\to\tau\nu)$ can be quite large
because of the chiral enhancement. If we assume, as a conservative
estimate, ${\rm Br}(B_c\to\tau\nu) < 10\%$, this translates to
\beq
\mid (a-b) y_2^\ast\mid  < 1.05 
\eeq
for $\Lambda = 1$ TeV.  Note that the limits one obtains from neutral
current mediated processes, like $\bsbsbar$ mixing or $B_s\to
\tau^+\tau^-$ \cite{bobeth}, are not valid in these cases as the
couplings are different.

\subsection{$B_s \to\tau^+\tau^-$}

The term from ${\cal O}_S$ that we would be interested in is
\beq
A\left[(\{x_1^\ast \bar{d} + x_2^\ast \bar{s} + x_3^\ast\bar{b}\}
P_R b)(\bar\tau P_L \tau) +
(\bar{b}P_L \{ x_1 d + x_2 s + x_3 b\})(\bar\tau P_R \tau)\right]\,,
\eeq
and the relevant operator is
\beq
A x_2^\ast (\bar{s}P_R b)(\bar\tau P_L\tau) = \frac12 A x_2^\ast \left[
(\bar{s} P_R b)(\bar\tau\tau) - (\bar{s} P_R b)(\bar\tau\gamma_5\tau)\right]\, .
\eeq
This gives
\beq
{\rm Br}(B_s\to\tau^+\tau^-) = \frac{G_F^2\alpha^2 m_{B_s}^5 f_{B_s}^2 \tau_{B_s}}{256\pi^3}
\sqrt{1-\frac{4m_\tau^2}{m_{B_s}^2}} \left[ \left(1-\frac{4m_\tau^2}{m_{B_s}^2}\right) {\cal A} + {\cal B}
\right]
\eeq
where
\bea
{\cal A}&=& \left\vert \frac{\zeta }{m_b+m_s}\right\vert^2\,,\nonumber\\
{\cal B}&=& \left\vert \frac{\zeta }{m_b+m_s} + \frac{2m_\tau}{m_{B_s}^2} [(V_{tb}V^\ast_{ts})C_{10} 
]\right\vert^2 \approx {\cal A}\,,\nonumber\\
\zeta&=& \frac{a x_2^\ast}{\Lambda^2} \frac{\sqrt{2}}{8G_F} \frac{4\pi}{\alpha}\,,
\eea
where $C_{10}$ is the Wilson coefficient for the corresponding SM
operator, and is too small to be of any consequence.  If we take the
upper limit of ${\rm Br}(B_s\to\tau^+\tau^-)$ to be 3.5\%, we get a bound, 
depending on definition used for $m_b$.
For example,
\beq
\vert a x_2^\ast\vert < 1.52 \, (1.34) \left( \frac{\Lambda}{1~{\rm TeV}}\right)^2
\left( \frac{{\rm Br}(B_s\to\tau^+\tau^-)}{3.5\%} \right)^{1/2}\,,
\eeq
for $m_b = m_b^{pole} = 4.8$ GeV ($m_b = m_b(m_b) = 4.2$ GeV). 
This agrees with Ref.~\cite{bobeth}.
Thus, potentially, $a$ can be large.

For the $\{V,A\}$ couplings coming from ${\cal O}_V$, one has
\beq
{\rm Br}(B_s\to\tau^+\tau^-) = 
\frac{f_{B_s}^2\tau_{B_s} m_\tau^2 m_{B_s}}{32\pi}  \sqrt{1-\frac{4m_\tau^2}{m_{B_s}^2}}
\left\vert \frac14 \frac{c}{\Lambda^2} x_2^\ast x_3\right\vert^2\,.
\eeq
Note that this is further suppressed by a factor of $m_\tau^2/m_{B_s}^2$, which results in a weaker
bound:
\beq
\vert cx_2^\ast x_3\vert < 6.0 \left( \frac{\Lambda}{1~{\rm TeV}}\right)^2
\left( \frac{{\rm Br}(B_s\to\tau^+\tau^-)}{3.5\%} \right)^{1/2}\,.
\eeq
\subsection{Width difference $\dgs$}

Following Ref.~\cite{bobeth}, let us quote the relevant expressions for the width difference
$\dgs$:
\beq
\begin{array}{rclcl}
{\cal O}_S &\Rightarrow& \dis \Gamma^s_{12, NP} & = & \dis 3 \, N \, x \, 
\sqrt{1-4x} \, \bra Q_S^R \ket \, C_S^2\,,\\
{\cal O}_V &\Rightarrow&
\dis \Gamma^s_{12, NP} & = & \dis N \, 
    \Big[ \left\{ 1 + (1-x) \, \sqrt{1-4x} \right\} 
          \, \bra Q_V^L\ket 
\\[1ex]
& & & & \dis \hspace*{1em} +
\left\{ 1 + (1+2x) \, \sqrt{1-4x} \right\} \, \bra Q_S^R \ket \Big] \, C_V^2\,,
\end{array}
\label{gamm12}
\eeq
where
\bea
x &=& m_\tau^2/m_{B_s}^2\,,\nonumber\\
 N &=& -\, \frac{G_F^2 m_b^2}{6\pi m_{B_s}} (V^\ast_{ts} V_{tb})^2\,,\nonumber\\
 \bra Q_S^R\ket &=& -\, \frac{5}{12} f_{B_s}^2 m_{B_s}^2 B_S\,,\nonumber\\
 \bra Q_V^L\ket &=& \frac23 f_{B_s}^2 m_{B_s}^2 B_V\,,\nonumber\\
 C_S &=& \frac{\sqrt{2}} {4G_F} \, \frac{1}{V^\ast_{ts}V_{tb}} \,
\frac{a x_2^\ast}{2 \, \Lambda^2}\,,\nonumber\\ 
 C_V &=& \frac{\sqrt{2}} {4G_F} \, \frac{1}{V^\ast_{ts}V_{tb}} \,
        \frac{c}{2\Lambda^2} x_2^\ast x_3\,.
 \eea
Note that the second equation of (\ref{gamm12}) has 
to be augmented by the inclusion of the 
$\nu_\tau$ loop, which can be obtained from
the corresponding $\tau$ contribution by putting $x=0$.  
For numerical evaluation, we use the lattice values $f_{B_s}=0.231$ GeV, $B_S =
1.3$, $B_V = 0.84$.
From $B\to K\tau\tau$, there is a (scale-independent) bound, namely $C_V < 0.8$, which translates to
\beq
\frac12 cx_2^\ast x_3 < 1.05 \left( \frac{\Lambda}{1~{\rm TeV}}\right)^2\,.
\eeq

\subsection{$\Delta M_s$ and the mixing phase $\phi_M$}

\begin{figure}
\vspace{-15pt}
\begin{center}
 \begin{picture}(55,120)(110.0,-30)
\ArrowLine(0,80)(40,40){psBlue}
    \Text(20, 75)[c]{$b(p_1)$}
\ArrowLine(40,40)(0,0){psBlue}
    \Text(20, 5)[c]{$s(p_2)$}
\ArrowArc(60,40)(20,0,180){psRed}
    \Text(60, 70)[c]{$\tau(k)$}
\ArrowArc(60,40)(20,180,360){psRed}
    \Text(60, 10)[c]{$\tau(k+q)$}
\ArrowLine(80,40)(120,80){psBlue}
    \Text(100, 75)[c]{$s(p_4)$}
\ArrowLine(120,0)(80,40){psBlue}
    \Text(100, 5)[c]{$b(p_3)$}

\ArrowLine(150,80)(190,40){psBlue}
    \Text(170, 75)[c]{$b(p_1)$}
\ArrowLine(190,40)(150,0){psBlue}
    \Text(170, 5)[c]{$s(p_4)$}
\ArrowArc(210,40)(20,0,180){psRed}
    \Text(210, 70)[c]{$\tau(k)$}
\ArrowArc(210,40)(20,180,360){psRed}
    \Text(210, 10)[c]{$\tau(k+q)$}
\ArrowLine(230,40)(270,80){psBlue}
    \Text(250, 75)[c]{$s(p_2)$}
\ArrowLine(270,0)(230,40){psBlue}
    \Text(250, 5)[c]{$b(p_3)$}

 \end{picture}
\end{center}
\vskip -40pt 
\caption{Typical one-loop corrections to $B_s$--$\bar B_s$ mixing
originating from four-fermion operators.}
\label{fig:BSBsbar}
\end{figure}
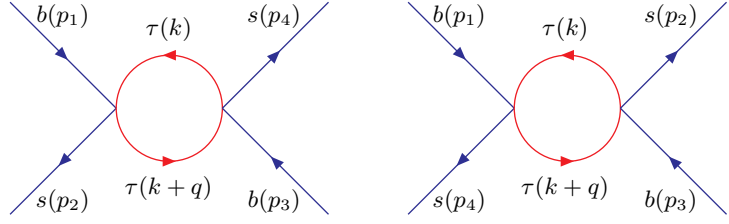

The aim of this subsection is to show how and why the constraints 
coming from $\Delta M_s$ measurements are so restrictive in nature.
Here, we will start from ${\cal O}_V$. There can be two sets of
possible diagrams, one with the $\tau$ lepton (see Fig.\ref{fig:BSBsbar})
and the other with the
neutrino. 
As the amplitudes are not chirality-suppressed, both the diagrams
contribute equally. The exact amplitudes cannot be calculated
unless we know about the ultraviolet completion of the effective
theory. 
If we use a cut-off regularization, the leading term, which is
divergent, should match with the leading term of the full theory. The
leading term of the loop amplitude is quadratically divergent, so we
can safely neglect the subleading terms.

The relevant part of the effective operator is
\beq
{\cal O}_V \supset \frac12 C (\bar{b}\gamma^\mu P_L b) (\bar\tau\gamma_\mu P_L \tau)
\to \frac12 Cx_2^\ast x_3 (\bar{s} \gamma^\mu P_L b) (\bar\tau\gamma_\mu P_L \tau)\,.
\eeq
This gives rise to a mixing amplitude
\bea
i M_{12s} &=& \left( \frac{1}{2\Lambda^2} c x_2^\ast x_3\right)^2 \bra O_1\ket (i\Gamma_2) \times 16\nonumber\\
&=&  \frac{i}{\pi^2} \left( cx_2^\ast x_3\right)^2 \Lambda^{-2} \bra O_1\ket \nonumber\\
&=& \frac{i}{3\pi^2} \left( cx_2^\ast x_3\right)^2 \Lambda^{-2} \eta_{B_s} M_{B_s} f_{B_s}^2 B_{B_s}\,,
\eea
where we have used
\beq
\Gamma_2 = \frac{\Lambda^2}{4\pi^2}\,,\ \ 
\bra O_1 \ket = \frac13 \eta_{B_s} M_{B_s} f_{B_s}^2 B_{B_s}\,,
\eeq
$\Gamma_2$ being the leading term of the loop amplitude, 
and $O_1 = [\bar{s}_\alpha\gamma^\mu(1-\gamma_5)b_\alpha][\bar{s}_\beta \gamma_\mu(1-\gamma_5) 
b_\beta]$, $\alpha$ and $\beta$ being colour indices. 
The factor of 16 can be understood in the following way: there is another crossed box, which, in an
effective theory, is something like a $t$-channel amplitude. This gives a factor of 2. The initial
meson can pick up a $\bar{b}$ from $O_1$ in two ways, and an $s$ in two ways, so the symmetry
factor is $4$. The neutrino mediated amplitude gives another factor of 2. 

\begin{figure}
\vspace{-15pt}
\centerline{
\rotatebox{0}{\epsfxsize=8.5cm\epsfysize=10.5cm\epsfbox{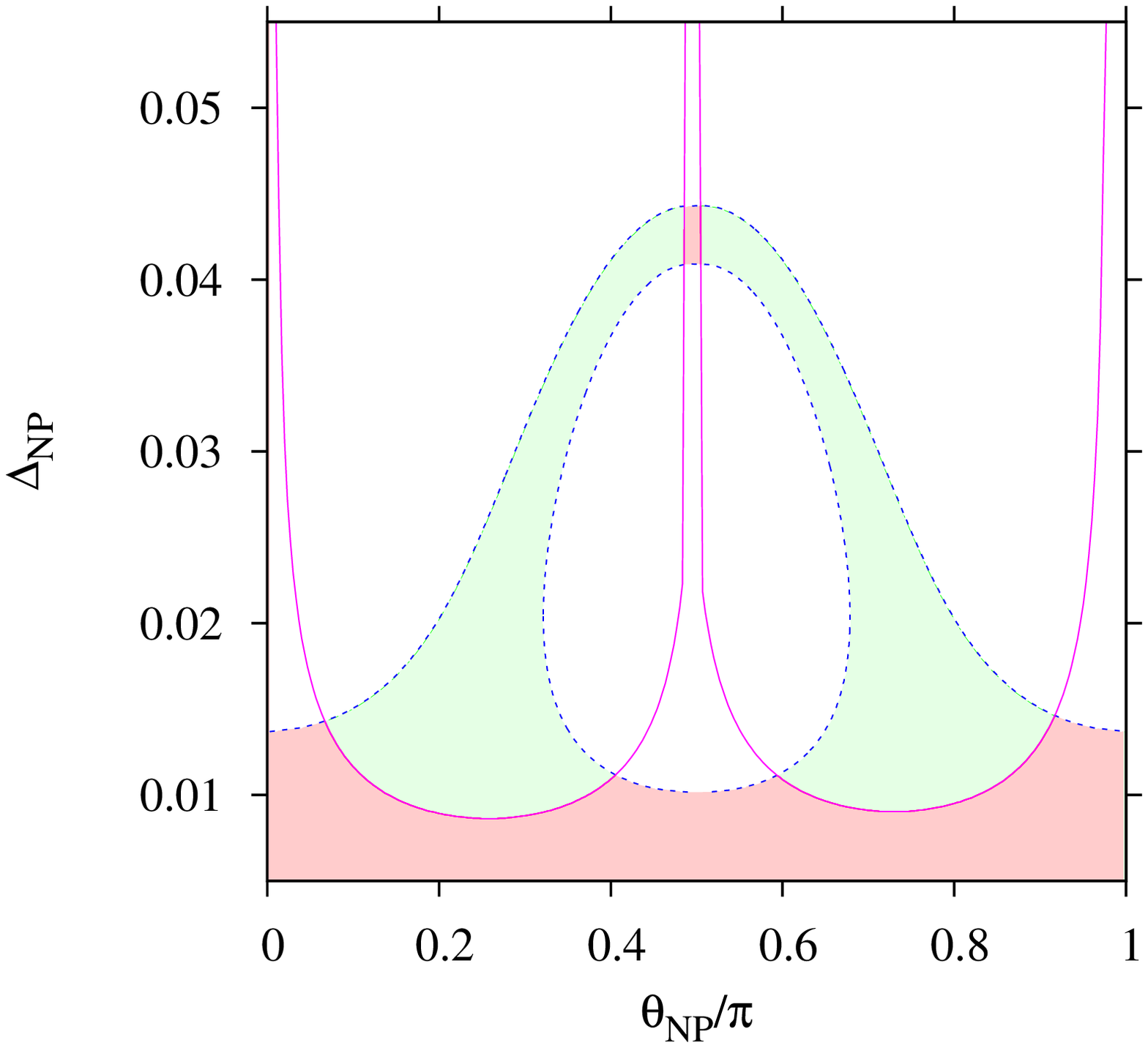}}
\rotatebox{0}{\epsfxsize=8.5cm\epsfysize=10.5cm\epsfbox{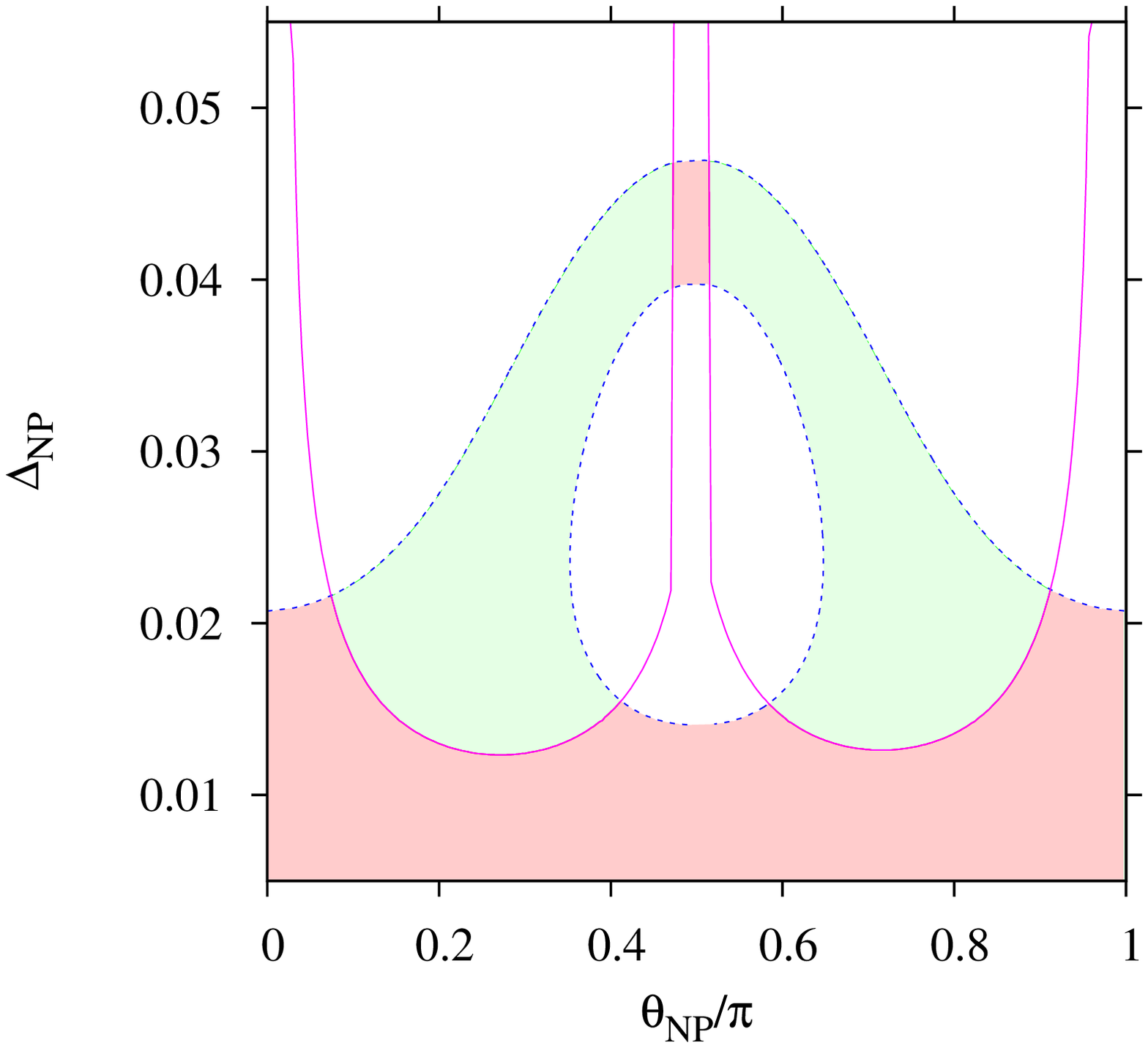}}
}
\vspace*{-18ex}
\caption{The allowed magnitude and phase of the coupling $cx_2^\ast
  x_3$ from the measurement of $\Delta M_s$ and $2\beta_s$, the
  effective mixing phase from the box amplitude. 
  {The light (green) shaded area between the dashed (blue) curves
    is allowed by $\Delta M_s$ measurement, whereas the area between the 
solid (purple) curves is allowed by the data on
  $2\beta_s$. Thus, only the two patches of dark (pink) shaded 
    area is finally allowed.}
  For the left (right)
  plot, the experimental errors are taken at $1(2)\sigma$ level. }
   \label{plotva01}
\end{figure}
Comparing with $i M_{12s}^{SM}$, we find
\beq
\frac{M^{NP}}{M^{SM}} = \frac{4(cx_2^\ast x_3)^2 \Lambda^{-2}} {G_F^2 m_W^2 (V_{tb}V_{ts}^\ast)^2 S_0(x_t)}\,,
\label{delm-sm-np}
\eeq
where $x_t=m_t^2/m_W^2$ and $S_0(x_t)$ is the relevant Inami-Lim function. 
The SM amplitude is GIM suppressed whereas there is no such suppression for the NP amplitude, and thus
Eq.\ (\ref{delm-sm-np}) puts a fairly tight constraint on $cx_2^\ast x_3$. If we want
the latter to be large, the phase
must be opposite to that of the SM amplitude, so that there is a destructive interference: $M^{NP}
\sim -2M^{SM}$. Taking the errors on the $\Delta M_s$ prediction in the SM and the measurement of the
same quantity both at $2\sigma$, we get
\beq
\left|cx_2^\ast x_3 \right|  < 0.048 \left( \frac{\Lambda}{1~{\rm TeV}}\right)\,.
\eeq
The allowed region is shown in Fig. \ref{plotva01}. For $\Lambda = 1$ TeV, the limits on $a^s_{fs}$ are
\beq
-6.3 (-11.2) \times 10^{-4} < a^s_{fs} < 2.2 (6.9) \times 10^{-4}
\eeq
at $1(2)\sigma$. 
Thus, by themselves, such operators are unable to explain the dimuon anomaly,
and the explanation must lie somewhere else. 
The maximum value of ${\rm Br}(B_s\to\tau^+\tau^-)$ is about $3\times 10^{-4}\%$.

The situation is marginally better for a chiral coupling in the scalar
sector; {\em i.e.}, either $S-P$ or $S+P$.  For such operators, the
leading term in the $\bsbsbar$ mixing amplitude is proportional to
$m_\tau^2\log\Lambda^4$, and there is no effective constraint from
$\Delta M_s$ and mixing phase. However, the major constraint comes
from ${\rm Br}(B_s\to\tau^+\tau^-)$, and also partially from
$\dgs$. We find
\beq
\vert a^s_{fs}\vert < 6\times 10^{-4}
\eeq
for ${\rm Br}(B_s\to\tau^+\tau^-) < 4\%$ and taking all the errors at $2\sigma$ level.
Thus, none of these schemes are enough to explain the dimuon anomaly completely.

\subsection{$H\to\tau^+\tau^-$}

Let us begin by parametrizing the tree-level Higgs-tau coupling by
\be
 {\cal L}_{\rm tree}^{(H \, \bar\tau \tau)} = h_\tau \, \bar\tau \, \tau \, H \ .
    \label{tree_Yukawa}
\ee
As in any quantum theory, this interaction Lagrangian receives 
quantum corrections. We neglect here all the SM corrections and concentrate 
solely on that wrought by the four-fermion operators.
The scalar-pseudoscalar operators give rise to an effective interaction of the form
\be
\frac{b}{2\Lambda^2}{\rm Re}(y_3) \left[(\bar{t}t)(\bar\tau\tau) - (\bar{t}\gamma_5 t) (\bar\tau\gamma_5\tau)\right]
+ 
\frac{b}{2\Lambda^2}{\rm Im}(y_3) \left[(\bar{t}t)(\bar\tau\gamma_5\tau) - (\bar{t}\gamma_5 t) (\bar\tau\tau)\right]\,.
\ee
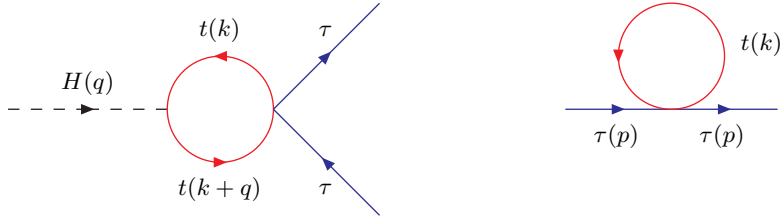
\begin{figure}
\vspace{-15pt}
\begin{center}
\begin{picture}(155,120)(55.0,-30)
\DashArrowLine(-20,40)(40,40){5}{psBlack}
    \Text(10, 50)[c]{$H(q)$}
\ArrowArc(60,40)(20,0,180){psRed}
    \Text(60, 70)[c]{$t(k)$}
\ArrowArc(60,40)(20,180,360){psRed}
    \Text(60, 10)[c]{$t(k+q)$}
\ArrowLine(80,40)(120,80){psBlue}
    \Text(100, 70)[c]{$\tau$}
\ArrowLine(120,0)(80,40){psBlue}
    \Text(100, 10)[c]{$\tau$}

\ArrowLine(190,40)(230,40){psBlue}
    \Text(210, 30)[c]{$\tau(p)$}
\ArrowLine(230,40)(270,40){psBlue}
    \Text(250, 30)[c]{$\tau(p)$}
\ArrowArc(230,60)(20,0,360){psRed}
    \Text(265, 65)[c]{$t(k)$}

\end{picture}
\end{center}
\vskip -50pt 
\caption{Typical one-loop corrections to 
(left) the $H \bar \tau \tau$ 
vertex and 
(right) $\tau$ self energy.}
\label{fig:feynman}
\end{figure}

Each of these terms generates, at one-loop, a two-point diagram
  contributing to the effective $H \bar \tau \tau$ coupling 
(see Fig.~\ref{fig:feynman}). The said
  diagrams are manifestly quadratically divergent and need to be
  regularized. Given that our basic theory is only an effective one,
  we may use a momentum cut-off regularization scheme, to yield the
  following correction to the Lagrangian of Eq.~(\ref{tree_Yukawa}):
\be  
\delta{\cal L}_{\rm 1-loop} = 
\frac{3bh_t}{8\pi^2} \, \frac{\Lambda_{\rm cutoff}^2}{\Lambda^2} 
   \, \left[ {\rm Re}(y_3) \; \bar\tau\tau 
           + {\rm Im}(y_3) \; \bar\tau\gamma_5\tau \right] \, H 
   + \cdots \, ,
\ee
where the ellipsis denote subleading terms. It is natural to
  consider $\Lambda_{\rm cutoff} = \Lambda$, for the two are expected
  to be similar. The appearance of a divergent correction to the
  pseudoscalar coupling (one that did not exist at the tree level)
  might seem disconcerting at first. However, it should be realised
  that we are dealing with a nonrenormalizable theory and the
  existence of such a divergence only reflects the fact that a large
  correction to $H \tau \tau$ is not prevented by the symmetries of
  the theory extant on admitting the general four-fermion
  interaction. On inclusion of the ultraviolet completion, 
  such divergences would disappear identically.
Clearly the two Lorentz structures contribute incoherently 
to $\Gamma(H \to \tau \tau)$. Formally though, the contribution of the 
scalar coupling correction may be larger 
as it can interfere with the 
SM amplitude, and thus can enter at an earlier order in the perturbation
theory. 
 For simplicity, though, let us assume that $y_3$ is real. Then,
we can parametrize the effective $H \tau \tau$ vertex, upto one-loop by
\be
\barr{rcl}
\dis {\cal L}_{\rm eff}^{(H \, \bar\tau \tau)} & = & \dis
    h_\tau \, (1 + \xi) \, \bar\tau \tau H + \cdots \\[2ex]
 \xi & = & \dis  \frac{3 b h_t y_3}{8 \pi^2 h_\tau}\,.
\earr
\ee
Similarly,
  several other Yukawa couplings also receive corrections, but these
  are suppressed on account of the particular structure of NP. For
example, the  bottom quark Yukawa coupling receives a correction 
on account of a tau-loop, and this change can
be expressed as $h_b \to h_b + b h_\tau y_3 / 8 \pi^2$.

\begin{figure}
\vspace{-70pt}
\centerline{
\rotatebox{0}{\epsfxsize=10cm\epsfbox{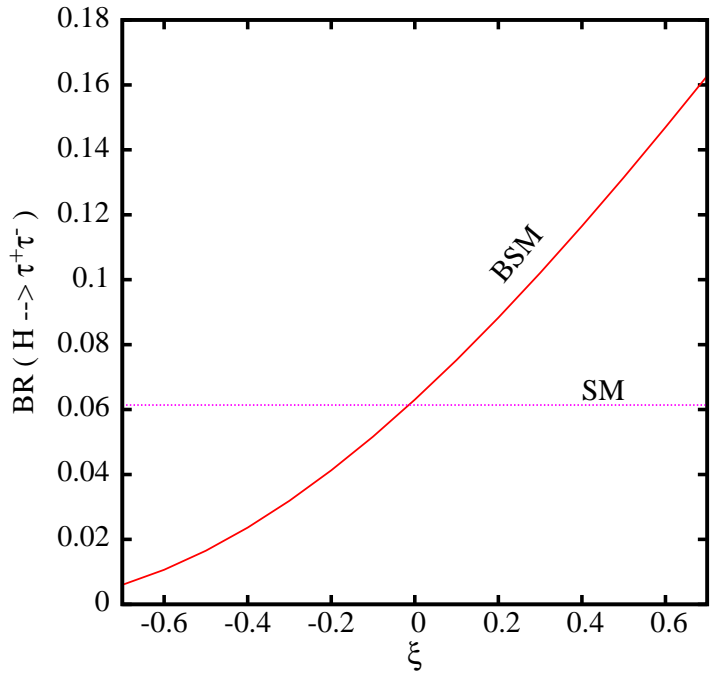}}
}
\vspace{-80pt}
\caption{The $H \to \tau^+\tau^-$ branching fraction as a function of $\xi$, 
which we take to be real for $m_H = 125 $ GeV. 
For comparison, we also show the SM branching ratio.}
   \label{xiBR}
\end{figure}
 
It might be argued that our calculation of $\xi$ is somewhat 
naive, and it is indeed true. However, an exact calculation 
necessitates a knowledge of the ultraviolet completion of the theory, 
and, in a sense, goes against the spirit of an effective theory. Nonetheless, 
$\xi$ does encapsulate the leading correction, and 
in Fig.\ \ref{xiBR}, we show the variation of the branching fraction of 
$H\to\tau^+\tau^-$ as a function of the real variable $\xi$ 
for $m_H = 125 $ GeV. 
From  Fig.\ \ref{xiBR}, it is very clear that even a moderate value of
 $\xi\sim -0.3$ is enough to give a 50\% suppression in the 
${\rm BR} (H \to \tau^+ \tau^-)$ . 
On the other hand, from the observed upper limit of 
$\sigma_{H\to \tau\tau}/\sigma_{SM} \approx 1.1$ from the LHC \cite{cms-higgs}, 
we get an upper limit of $\xi\approx 0.05$. 
 
For the $\{V,A\}$ current, 
once again, both scalar and pseudoscalar couplings appear 
at one-loop. However, the loop is convergent as the current structure 
demands that extra powers of the fermion masses must be picked up. Hence,
the corresponding corrections are too small to be of any consequence.

\subsection{Anomalous top decays}

One might ask whether the new couplings will lead to observable
rates for FCNC top decays, {\em e.g.} $t\to c\tau^+\tau^-$. Unfortunately 
though, even if we use values of the
couplings $b \, y_2^\ast$ (for scalar operators) or $c \, y_2^\ast$
(for vector operators) significantly larger than what we need to
explain the anomalies under investigation, the rates for this decay
are still much smaller than the LHC sensitivity limits. For example,
we might naively use the limit on the branching ratio of $t\to
cZ$~\cite{CMS_top_fcnc}, namely 
\be
{\rm Br}(t \to c Z) < 0.24\%, 
    \label{t_cZ}
\ee
along with
the measurement \cite{D0_topwidth} of the decay width of the top
\be 
\Gamma_t = 2.00^{+0.47}_{-0.43}~{\rm GeV}
\ee
to yield the very weak limit of 
$\Lambda \gtap 0.5 \, m_t$ when the couplings are only 
restricted to be perturbative. Furthermore, even the use of Eq.(\ref{t_cZ})
is over-optimistic, for the CMS limits have been derived requiring that 
the $Z$-mass can be reconstructed from its decay products. In the current case,
this does not apply and the signal to noise ratio is lower than that assumed
to obtain Eq.(\ref{t_cZ}). In other words, the actual limit is much weaker 
than that quoted above.

\section{Conclusions}

In this paper, we have investigated
the possible implications of a scenario that
involves some new interactions involving the third generation
fields. Any model that treats the third generation differently from
the first two generations may lead to such a scenario. 
Without attempting to prescribe an ultraviolet-complete theory,
  we rather consider an effective theory valid below some cutoff
scale $\Lambda$, above which the full theory takes over. A possible
motivation for such a scenario is the fact that there are
excesses over the SM predictions for the charged current B-decays, namely $B
\to D(D^\ast) \tau\nu$ and $B^+\to\tau\nu$, while the predictions for
the processes involving the first two generations of leptons do not
show any tension with the data.

In the effective theory, there can be several four-fermion operators
involving the third generation fields and 
several possible choices for the Lorentz structures
of the currents. 
With the Wilson coefficients for tensor operators being 
severely constrained by the data on radiative decays, we preclude 
these from our discussions.
With $\Lambda$ being larger than the electroweak scale, 
it is quite likely that such four-fermion operators in the effective Lagrangian
should be written in the weak basis, and for reasons of economy, 
we consider only one such operator at a time.
Rotating the fields  to the mass basis generates new
operators involving first and second generation quark fields, 
albeit suppressed by the
corresponding entries of the quark mixing matrix.

Once we have a set of such operators, we study their implications on
several B-decay observables. In particular, we show that the apparent
excesses in the B-decay channels mentioned above can be accommodated
satisfactorily in this scenario;
complementary observables lead to nontrivial constraints
on the model parameters. The vector-axial vector operators
successfully explain the excesses in both $B\to D\tau\nu$ and $B\to
D^\ast \tau\nu$ channels, 
apart from leading to a sizable enhancement to the
$B_c\to\tau\nu$ branching ratio as a testable prediction. The
scalar-pseudoscalar couplings are not that successful in explaining
both the excesses, but there is a definite improvement over the SM
predictions.  The excess in the channel $B^+\to\tau\nu$ can have a
satisfactory explanation too, although the tension is no longer worrying.

The operators leading to $\bdbdbar$ and $\bsbsbar$ mixing are more
constrained. They have identical Lorentz structures as those discussed
before, but with different quark fields and different Wilson
coefficients. While these coefficients are 
constrained from the 
measured mass
differences $\Delta M_d$ and $\Delta M_s$, 
the restrictions are not strong enough
 to rule out any observable enhancement in the
$B_s\to\tau^+\tau^-$ channel, which should be investigated more
carefully as one of the best windows to new physics.
Unfortunately though, the
anomalously large dimuon asymmetry receives only a marginal
improvement over the SM prediction. One might need other operators to
explain this, but it is not easy given the tight constraints from
$\Delta M_s$ measurements.

One thing that still remains unobservably small in this class of
models is anomalous top decay, like $t\to c\tau^+\tau^-$. The other
side of the coin is that if such decays are observed, the new physics
must be something different from those described here, as the
expectations will be in conflict with the $B$-decay observables.

It is not yet certain whether there is a deficiency in the
$H\to\tau^+\tau^-$ channel, but at the $1\sigma$ level, the
cross-section is slightly below the SM prediction. While it is too
early to say anything about this channel, we would like to point out
that the interactions discussed in this paper can potentially modify
the predictions for this channel, without disturbing those for other
channels.  Further data from LHC will be eagerly anticipated.


\acknowledgments

We acknowledge Swagoto Banerjee for illuminating discussions,
and Diptimoy Ghosh for bringing the latest Belle result on $B\to
\tau\nu$ to our notice. 
The work of AK was supported by CSIR, Government of India,
and the DRS programme of the University Grants Commission.
Both DC and DKG would like to thank the High Energy Physics 
Group of ICTP for hospitality where this project was started.
DKG would also like to acknowledge the hospitality provided by the
University of Helsinki and the Helsinki Institute of Physics
where part of this work was done.


\end{document}